\documentclass[
preprint,
showpacs,
showkeys,
nofootinbib,
aps,
amsfonts,
amssymb]{revtex4-1}

\usepackage{graphicx} 
\usepackage{amsmath, amssymb, bm} % AMS packages 
\usepackage[dvips]{color}

\usepackage{psfrag,epsfig}
\usepackage{natbib}
\usepackage{hyperref}

\usepackage{csvsimple} %Load CSV file as tables
\usepackage{adjustbox} %  Fit wide table

\usepackage{ifthen}

\def\){\right)} 
\def\({\left(} 
\def\]{\right]} 
\def\[{\left[}

\def\nLi{$^7\mathrm{Li}(n, \gamma)^8\mathrm{Li}$}

\def\HeAlpha{$^3\mathrm{He}(\alpha,\gamma)^7\mathrm{Be}$}

\def\P {\left(\frac{\stackrel{\rightarrow}{\nabla}}{m_\phi}-\frac{\stackrel{\leftarrow}\nabla}{m_\psi}\right)}

\def\erasepar#1{}

\definecolor{darkred}{rgb}{0.75,0.1,0.1}
\definecolor{darkblue}{cmyk}{1,1,0,0.35}
\definecolor{brown}{rgb}{0.55, 0.05, 0.05}
\definecolor{paleblue}{cmyk}{0.7,0.7,0.0,0.0}
\definecolor{darkgreen}{cmyk}{1,0,1,0.5}

\begin{document}

\title{
Radiative \HeAlpha ~reaction 
in Halo Effective Field Theory}

\author{%
Renato Higa $^{a}$}
\email{higa@if.usp.br}

\author{%
Gautam Rupak $^b$}
\email{grupak@u.washington.edu}

\author{%
Akshay Vaghani $^b$}
\email{av298@msstate.edu}

\affiliation{ $^a$ Instituto de F\'\i sica, Universidade de S\~ao Paulo, 
R. do Mat\~{a}o 1371, 05508-090, S\~ao Paulo, SP, Brazil\\
$^b$ Department of Physics $\&$ Astronomy and 
HPC$^2$ Center for Computational Sciences, 
Mississippi State
University, Mississippi State, MS 39762, U.S.A.}

\begin{abstract}
In this work we study the radiative capture of ${\rm {}^3He}$ on 
${\rm {}^4He}$ within the halo effective field theory (EFT) framework. 
At leading order the capture amplitude comprises the initial state 
$s$-wave strong and Coulomb interactions summed to all orders.
At the same order in the expansion, leading two-body currents contribute 
as well. We find delicate cancelations between the various contributions, 
and the two-body current contributions can be replaced by appropriately 
enhancing the asymptotic normalizations of the $^7$Be ground and first 
excited state wave functions. 
The next-to-leading order corrections come from the $s$-wave shape parameter 
and the pure Coulomb $d$-wave initial state interactions. 
We fit the EFT parameters to 
available scattering data and most recent capture data. 
Our zero-energy astrophysical 
$S$-factor estimate, $S_{34}\sim 0.55$ keV b, is consistent within 
error bars with the average in the literature. 
\end{abstract}


\preprint{INT-PUB-16-055}
\maketitle

%===========================================================
\section{Introduction}\label{sec_intro}
%===========================================================
%\input{sec1-intro}
Low-energy reaction rates involving light nuclei have become a recurrent 
subject nowadays, given their importance in many astrophysical processes. 
As astronomical observations aim at more accuracy, comparable improvements 
from experiments and theoretical estimates are desired for these reactions. 
An example is the process 
${\rm {}^3He(\alpha,\gamma)^7Be}$ that takes place in the interior of our 
Sun. The astrophysical $S$-factor for this reaction within the Gamov window 
$E_G\sim 20$ keV is the main source of uncertainty in the solar neutrino flux 
detected on Earth. For instance, the flux of neutrinos from the $\beta^+$ 
decay of ${\rm {}^8B}$ and from the electron capture on ${\rm {}^7Be}$ 
are proportional to $[S_{34}]^{0.81}$ and $[S_{34}]^{0.86}$, 
respectively~\cite{Cyburt:2008up,Bahcall:1987jc,Adelberger:1998qm}. 
The first weak decay provides energetic solar neutrinos that were 
detected by Super-K~\cite{Fukuda:2001nj} and SNO~\cite{Aharmim:2006kv} but 
depends also on the ${\rm {}^7Be(p,\gamma)^8B}$ reaction rate. 
Electron capture on ${\rm {}^7Be}$, on the other hand, provides a solar 
neutrino flux three orders of magnitude higher than the former 
process~\cite{Bahcall:2004mq,Bahcall:2004pz}, 
with less energetic neutrinos which can be measured by the BOREXINO 
experiment at Gran Sasso~\cite{Arpesella:2008mt}, and depends 
exclusively on the ${\rm {}^3He(\alpha,\gamma)^7Be}$ reaction.
In either case, the need for a better description of the latter at very low 
energies is of prime importance in order to improve constraints 
from solar neutrinos, like mass hierarchy, flavor mixing angles and CP 
violating phases. 
Besides neutrino physics, the ${\rm {}^3He(\alpha,\gamma)^7Be(e^-,\nu)^7Li}$ 
chain reaction is the main source of ${\rm {}^7Li}$ production during big bang 
nucleosynthesis (BBN), with a Gamov window 
$100\;{\rm keV}\lesssim E_G\lesssim 900\;{\rm keV}$. 
The primordial abundance of 
${\rm {}^7Li}$ calculated from BBN and WMAP cosmic baryon 
density measurements is a factor of 3 to 4 times larger than observations 
of metal-poor stars in our 
galaxy~\cite{Cyburt:2008up,Cyburt:2004cq,Fields:2011zzb}, which constitutes 
the so-called lithium problem. 
Many proposals to solve this puzzle, that involves alternative astronomical 
measurements and modeling, nuclear, and particle physics, can be strongly 
constrained with more reliable information on $S_{34}$, since the 
abundance ratio 
${\rm {}^7Li/H}\propto [S_{34}]^{0.96}$~\cite{Cyburt:2008up,Cyburt:2004cq}. 

Given its importance to the topics mentioned above, several measurements of 
the ${\rm {}^3He(\alpha,\gamma)^7Be}$ reaction were done in the past 
(see~\cite{Adelberger:1998qm} and references therein) and most recent 
years~\cite{Cyburt:2008up,Adelberger:2010qa}. 
As pointed out in Ref.~\cite{Cyburt:2008up}, measurements done prior to the 
review article~\cite{Adelberger:1998qm} fall into two discrepant groups---those based 
on induced ${\rm {}^7Be}$ activity, and those relying on prompt $\gamma$-ray 
detection. Due to improvements in detectors and background suppressions, 
this discrepancy is no longer present in the most recent 
measurements~\cite{Adelberger:2010qa}. 
Nevertheless, error bars are still relatively large in the low-energy regime 
of astrophysical interest, due to the strong suppression of events by the 
Coulomb repulsion. 
The higher energy data where statistics are better shall therefore be 
theoretically extrapolated down to astrophysically relevant energies in 
an as less model-dependent way as possible. 

The $^7$Be nucleus has a predominant $^3$He-$\alpha$ cluster structure. 
Its ground state binding energy, $B_0\sim 1.6$~MeV, is considerably smaller 
than the proton separation energy in $^3$He ($S_p\sim 5.5$~MeV) and the 
energy of the first excited state of the $\alpha$ particle ($\sim 20$~MeV). 
The distinct two-cluster configuration of $^7$Be, with tight constituents 
and the low-energy 
regime one is interested in, make this reaction very suitable for a 
halo effective field theory (halo EFT) approach. 
Halo EFT was first formulated in Refs.~\cite{Bertulani:2002sz,Bedaque:2003wa} 
in their study of the shallow $p$-wave neutron-alpha resonance and applied 
to other systems, such as the $s$-wave alpha-alpha 
resonance~\cite{Higa:2008dn,Gelman:2009be}, three-body halo 
nuclei~\cite{Canham:2008jd,Canham:2009xg}, coupled-channel 
proton-$^7{\rm Li}$ scattering~\cite{Lensky:2011he}, electromagnetic 
transitions~\cite{Hammer:2011ye} 
and capture reactions~\cite{Rupak:2011nk,Fernando:2011ts,Rupak:2012cr,
Zhang:2013kja,Zhang:2014zsa,Ryberg:2014exa,Zhang:2015ajn,Rupak:2016mmz}. 
In this work, 
we apply the same ideas to the ${\rm {}^3He(\alpha,\gamma)^7Be}$ radiative 
reaction, following a two-cluster approach of point-like objects at
leading order (LO) 
approximation. Corrections due to the structure of each cluster and 
higher order electromagnetic interactions are taken into account 
in perturbation theory. 
In halo EFT, a systematic and model-independent 
expansion of observables is achieved through the use of an expansion 
parameter ---formed by the ratio of a soft momentum scale $Q$, associated 
with the shallowness of the binding of the clusters, and a hard momentum scale 
$\Lambda$, related to the tightness of the cores. 
Moreover, the formalism guarantees unambiguous inclusion of electromagnetic 
interactions that preserve the required symmetry constraints, 
such as gauge invariance.

The paper is organized as follows. In Sec.~\ref{sec_lagrangian} we briefly 
comment on the energy scales, degrees of freedom, and channels relevant to 
the dominant E1 transition, as well as the construction of the corresponding 
interaction lagrangian. Sec.~\ref{sec_coulomb} presents the main elements 
necessary to deal with Coulomb interactions between the $^3$He and $\alpha$ 
nuclei. 
The amplitude for both initial elastic scattering state and final bound state  
are obtained in the halo EFT framework in 
Sec.~\ref{sec_elastic}.
 There we also relate the EFT couplings to the 
effective range parameters and set the power-counting. 
Sec.~\ref{sec_capture} collects the relevant expressions for the 
capture amplitude and cross section, whose numerical results are shown 
and discussed in Sec.~\ref{sec_analysis}. 
We present the EFT power-counting here. Our concluding remarks are 
presented in Sec.~\ref{sec_summary}.

%===========================================================
\section{Interaction}\label{sec_lagrangian}
%===========================================================
%\input{sec2-interaction} 

The halo EFT we construct treats the $^7$Be nucleus as a bound state of 
point-like nuclear clusters $^3$He and $\alpha$. The $\frac{3}{2}^-$ ground 
state has a binding energy $B_0=1.5866$ MeV, and the $\frac{1}{2}^-$ first 
excited state has a binding energy $B_1=1.1575$ MeV. The next excited 
state of $^7$Be is about 3 MeV above the $^3$He-$\alpha$ 
threshold~\cite{tunl}. In halo 
EFT the ground and first excited states are included respectively as 
$^2P_{3/2}$ and $^2P_{1/2}$ in the spectroscopic notation $^{2S+1}L_J$. 
The states beyond the first excited state are not included in the low-energy 
theory. Similarly, only the ground states of $^3$He and $\alpha$ are 
relevant at astrophysical energies. 

Early works on  radiative capture \HeAlpha ~indicate it is dominated by the E1 
transition from the initial $s$-wave state at low energies, see Ref.~\cite{Cyburt:2008up}. 
In the phase shift analysis of experimental data we use, the Coulomb subtracted $d$-wave phase shift is found to be small and treated as zero~\cite{Boykin:1972}. In the EFT we include initial $d$-wave state with only Coulomb but no strong interaction. 
Thus we consider 
the following Lagrangian for the calculation, 
%-----------------  Equations 
\begin{align}
\mathcal L=&\psi^\dagger\left[i\partial_0+\frac{\nabla^2}{2 m_\psi}\right]\psi 
+\phi^\dagger\left[i\partial_0+\frac{\nabla^2}{2 m_\phi}\right]\phi\nonumber\\
&+{\chi_{[j]}^{(\zeta)}}^\dagger\left[\Delta^{(\zeta)}+i\partial_0
+\frac{\nabla^2}{2M}\right]\chi_{[j]}^{(\zeta)}
+h^{(\zeta)}
\left[
{\chi_{[j]}^{(\zeta)}}^\dagger \psi P_{[j]}^{(\zeta)}\phi+\operatorname{h.c.}
\right],
\end{align}
where the spin-1/2 fermion field $\psi$
represents the $\frac{1}{2}^+$ ${}^3$He nucleus field with mass 
$m_\psi=2809.41$ MeV, and the scalar field
$\phi$ represents the spinless $0^+$ $\alpha$ 
field, with mass $m_\phi= 3728.4$ MeV. $M=m_\psi+m_\phi$ is the total mass.
We use natural units with $\hbar=1=c$. 
Note that, for some physical quantities, we keep more significant digits 
than necessary until presenting our final results. 
The projectors $P_{[j]}$ and the auxiliary fields $\chi^{(\zeta)}_{[j]}$ 
carry vector and spinor indices $[j]$ to specify the relevant spin-angular 
momentum channels for the incoming 
states $\zeta={}^2S_{1/2}$, ${}^2D_{3/2}$, ${}^2D_{5/2}$, 
final ground 
state $\zeta={}^2P_{3/2}$, and final excited state $\zeta={}^2P_{1/2}$, 
described below. We use the shorthand notation $\zeta=\pm$ to refer to the $^2P_{3/2}$ and $^2P_{1/2}$
channels, respectively. 

The $s$-wave interaction can be written using a spin-1/2 auxiliary field 
$\chi^{\alpha, s}$ as:
%-----------------  Equations 
\begin{align}
{\chi^{\alpha, s}}^\dagger\left[\Delta^{(s)}+i\partial_0
+\frac{\nabla^2}{2M}\right]\chi^{\alpha, s}
+h^{(s)}\[{\chi^{\alpha,s}}^\dagger P^{\alpha\beta,s}\psi^\beta\phi
+\operatorname{h.c.}\],
\end{align}
where the spinor indices $\alpha$, $\beta$ on the fields
$\chi^{\alpha,s}$ and $\psi^\beta$ 
are contracted using the diagonal $s$-wave projector 
$P^{\alpha\beta,s}=\delta^{\alpha\beta}$. The spinor index $\alpha=1,2$. 
The two $s$-wave couplings 
$\Delta^{(s)}$ and $h^{(s)}$ can be fitted to scattering length $a_0$ and 
effective range $r_0$ for elastic scattering of $^3$He and $\alpha$. 
We discuss this in more detail when we consider the relevant power-counting. 

For the final state, we want to project the vector index $i=1$, 2, 3 for 
the $p$-wave and the spinor index $\alpha$ for the $^3$He spin into 
total angular momentum $j=1/2$ and $j=3/2$ pieces. This can be done for a 
generic auxiliary field $\chi_i^\alpha$ as follows:
%-----------------  Equations 
\begin{align}
\chi_i^\alpha=\frac{1}{3}(\sigma_i\sigma_j)^{\alpha\beta}\chi_j^\beta
+\[ \delta_{ij}\delta^{\alpha\beta} 
-\frac{1}{3}(\sigma_i\sigma_j)^{\alpha\beta}\]\chi_j^\beta\,,
\end{align}
where the two pieces are the irreducible forms representing the $^2P_{1/2}$ 
and $^2P_{3/2}$ channels, respectively. The Pauli matrices $\sigma_i$'s act on 
the spinor indices. The two $p$-wave interactions can then be written as 
%-----------------  Equations 
\begin{align}
{\chi^{\alpha,\zeta}_i}^\dagger\left[\Delta^{(\zeta)}+i\partial_0
+\frac{\nabla^2}{2M}\right]\chi^{\alpha,\zeta}_i+
\sqrt{3}h^{(\zeta)}\[{\chi^{\alpha,\zeta}_i}^\dagger P_{ij}^{\alpha\beta,\zeta}\psi^\beta\P_j\phi+\operatorname{h.c.}\],
\end{align}
where the $p$-wave projectors are
%-----------------  Equations 
\begin{align}
P_{ij}^{\alpha\beta,\zeta}&= \frac{1}{3}(\sigma_i\sigma_j)^{\alpha\beta}\qquad 
\mathrm{for}\ \zeta={}^2P_{1/2}\,,\nonumber\\
P_{ij}^{\alpha\beta,\zeta}&
=\delta_{ij}\delta^{\alpha\beta} -\frac{1}{3}(\sigma_i\sigma_j)^{\alpha\beta}
\qquad \mathrm{for}\ \zeta={}^2P_{3/2}\,.
\end{align}
The two couplings $\Delta^{(\zeta)}$, $h^{(\zeta)}$ in each of the two 
$p$-wave channels can be determined from the corresponding binding momentum 
and effective range. For bound states, both of these couplings contribute at 
LO~\cite{Bertulani:2002sz,Bedaque:2003wa}. This remains true even 
in the presence of long-range Coulomb interaction as we have here. 

The capture calculation proceeds through the E1 transition. 
For one-body currents we couple the 
external photon through minimal substitution, that corresponds to gauging 
the momentum of the charged particle, $\bm{p}\rightarrow\bm{p}+Z e \bm{A}$, 
where $Z$ is the charge number. 
We include the long-range Coulomb interaction between the 
$^3$He and $\alpha$ nuclei  to all orders in perturbation by summing the 
Coulomb ladder as described below. Two-body currents that are not
related to elastic scattering operators by gauge invariance also contribute 
to the E1 transition between $s$-wave, 
and $p$-wave ground and excited states. These can be written using the auxiliary 
fields as 
%-----------  Equation
\begin{align}\label{eq:TwoBodyCurrentL1}
e \mu\( \frac{Z_\phi}{m_\phi}-\frac{Z_\psi}{m_\psi}\)L^{(\zeta)}_\mathrm{E1}\sqrt{3}h^{(s)} h^{(\zeta)}
{\chi^{\alpha,\zeta}_i}^\dagger P_{ij}^{\alpha\beta,\zeta}\chi^{\beta,s} E_j\,,
\end{align}
where $\mu$ is the reduced mass, $\bm{E}$ is the electric field, $Z_\psi=2$ 
and  $Z_\phi=2$ are the charge numbers of $^3$He and $\alpha$, respectively. 
We include factors of $h^{(s)}$, $h^{(\zeta)}$ and the effective charge 
$e\mu(Z_{\phi}/m_{\phi}-Z_{\psi}/m_{\psi})$
in the definition of the 
coupling $L_{E1}$. In the absence of Coulomb 
interaction, for a single charged particle, this reduces to a factor of 
$\sim 2\pi/\Big(\mu\sqrt{r_0 r_1^{(\zeta)}}\Big)$ 
that has been suggested earlier~\cite{Beane:2000fi}. 
Two-body currents such as this are usually not included in 
potential model calculations. 
Note that in our halo EFT formalism the one- and two-body currents are 
effective ones, whose origins lie in a more complicated one- and many-nucleon 
electromagnetic currents. The intricate contributions between different 
many-nucleon currents are discussed in a variational 4-nucleon study of 
Refs.~\cite{Carlson:1990nh,Carlson:1991ju} and in a three-cluster 
($p$-$d$-$^4$He) model of Ref.~\cite{Sadeghi:2013nba}, but is beyond 
the scope of this work.

%===========================================================
\section{Coulomb ladder}\label{sec_coulomb}
\begin{figure}[tbh]
\begin{center}
\includegraphics[width=0.68\textwidth,clip=true]{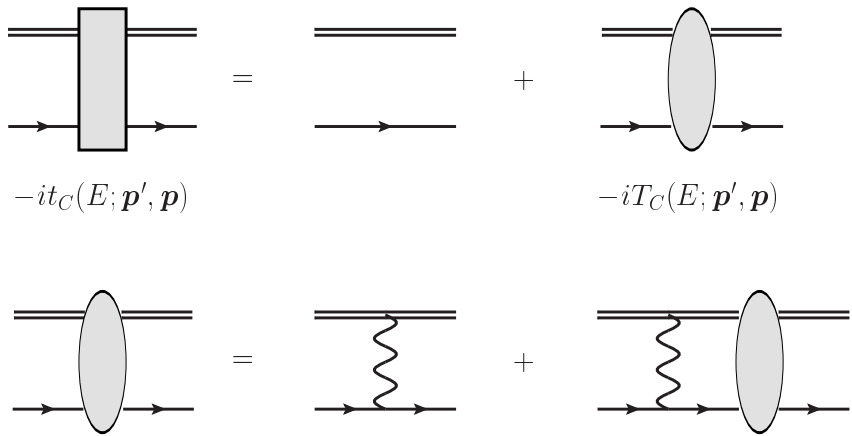} 
\end{center}
\caption{\protect Coulomb ladder diagrams. The double line represents the 
scalar
$\alpha$ particle, and the single line with an arrow the fermionic
$^3$He nucleus. The wavy lines 
represent photons. We only include Coulomb photon interaction between the 
charged particles.}
\label{fig:CoulombLadder}
\end{figure}

%===========================================================
%\input{sec3-coulomb}
For the scattering of two charged particles $^3$He and $\alpha$ at low energy, 
the relevant quantity that provides the strength of Coulomb photon exchanges 
is the Sommerfeld parameter $\eta_p= \alpha_e Z_\psi Z_\phi\mu/p$, where 
$\alpha_e=e^2/(4\pi)\approx 1/137$ is the electromagnetic fine structure 
constant, and $p$ is the relative center-of-mass 
(c.m.) momentum. The inverse of the Bohr radius of the system defines the 
momentum scale $k_C=\alpha_{e}Z_{\psi}Z_{\phi}\mu$ and the Sommerfeld 
parameter is written as the ratio $\eta_p=k_C/p$. Each photon exchange is 
proportional to $\eta_p$. In the low-energy
region that we consider, $p\lesssim k_C$, multiple photon exchanges 
contribute at least at the same order, forcing the summation of Coulomb 
ladder diagrams, Fig.~\ref{fig:CoulombLadder}.
%%%------------- Figure  1 --------------------------------

The Coulomb scattering amplitude $T_C$ satisfies the following useful 
relations 
%----------------- Equation 
\begin{align}
|\chi_p^{(\pm)}\rangle &=|\bm{p}\rangle
+\widehat G_0^{(\pm)}\widehat T_C|\bm{p}\rangle\,,\nonumber\\
\widehat G_C^{(\pm)}&=\widehat G_0^{(\pm)}+\widehat G_0^{(\pm)} 
\widehat T_C \widehat G_0^{(\pm)}\,,
\end{align}
where $|\bm{p}\rangle$ is a plane wave state in the two-particle 
c.m. system, and $|\chi^{(+)}_k\rangle$ 
($|\chi^{(-)}_k\rangle$) is the incoming (outgoing) Coulomb scattering state. 
The free and Coulomb Green's functions as 
operators are respectively written as 
%-------------  Equation
\begin{align}
&\widehat G_0^{(\pm)}=(E-\widehat H_0\pm i\epsilon)^{-1}\,,\nonumber\\
&\widehat G_C^{(\pm)}=(E-\widehat H_0-\widehat V_C\pm i\epsilon)^{-1}\, .
\label{eq:GreenFunction}
\end{align}
The $\pm i\epsilon$ signs in the definitions correspond to retarded and 
advanced Green's functions. Taking expectation values between final 
$\langle\bm{p}'|$ and initial $|\bm{p}\rangle$ momentum states, one can 
derive several useful relations, 
%-------------- Equation 
\begin{align}
G_0^{(\pm)}(E;\bm{p}',\bm{p})&
\equiv\langle\bm{p}'|\widehat G_0^{(\pm)}|\bm{p}\rangle
=\frac{(2\pi)^3\delta(\bm{p}'-\bm{p})}{E-p^{\prime\,2}/(2\mu)\pm i\epsilon}\,,
\nonumber\\
\frac{T_C(E;\bm{p}',\bm{p})}{E-p^{\prime\,2}/(2\mu)\pm i\epsilon} &
\equiv\langle\bm{p}'|\widehat G_0 \widehat T_C|\bm{p}\rangle
=
\chi_p^{(\pm)}(\bm{p}')-(2\pi)^3\delta(\bm{p}'-\bm{p})\,,\nonumber\\
G_C^{(\pm)}(E;\bm{p}',\bm{p})&\equiv
\langle\bm{p}'|\widehat G_C^{(\pm)}|\bm{p}\rangle
\nonumber\\&=
G_0^{(\pm)}(E;\bm{p}',\bm{p})
+\frac{T_C(E;\bm{p}',\bm{p})}{\left[E-p^{\prime\,2}/(2\mu)\pm i\epsilon\right]
\left[E-p^2/(2\mu)\pm i\epsilon\right]}\,, 
\end{align}
with $\chi_p^{(\pm)}(\bm{p}')\equiv\langle\bm{p}'|\chi_p^{(\pm)}\rangle$. 
The diagrammatic relation between $\widehat t_C$ and $\widehat T_C$ from 
Fig.~\ref{fig:CoulombLadder} can be expressed as 
%------------- Equation
\begin{align}
t_C(E;\bm{p}',\bm{p})&=(2\pi)^3\delta(\bm{p}'-\bm{p}) 
\(E -\frac{p^{\prime\,2}}{2\mu}\pm i\epsilon\)
+T_C(E;\bm{p}',\bm{p})\,,
\end{align}
and helps write $t_C(E;\bm{p}',\bm{p})$ in terms of $\chi_p^{(\pm)}(\bm{p}')$ 
and $G_C^{(\pm)}(E;\bm{p}',\bm{p})$. 

The Coulomb wave function and the retarded Green's function are known in 
closed form in coordinate space~\cite{Abramowitz,nist}, 
%------------ Equation
\begin{align}
\tilde\chi_{p}^{(\pm)}(\bm{r})&\equiv
e^{-\frac{\eta_p\pi}{2}}\Gamma(1\pm i\eta_p)\;
{}_1F_1(\mp i\eta_p,1;\pm ipr-i\bm{p}\cdot\bm{r})e^{i\bm{p}\cdot\bm{r}}\,,\nonumber
\\
\tilde\chi_{p}^{(+)}(\bm{r})&=\sum_{l=0}^{\infty}(2l+1)i^{l}e^{i\sigma_{l}}
P_{l}(\hat{\bm{p}}\cdot\hat{\bm{r}})\frac{F_{l}(\eta_p,pr)}{pr}\,,
\nonumber\\
G_C^{(+)}(E;\bm{r}',\bm{r})&\equiv\sum_{l=0}^\infty (2l+1)
P_l (\hat{\bm{r}}'\cdot \hat{\bm{r}})G^{(l)}_C(E, r',r)\,,\nonumber\\
G^{(l)}_C(E, r',r)&=-\frac{\mu p}{2\pi} \frac{F_l(\eta_p,r_{<} p)}{r_{<} p}
\frac{H^{(+)}_l(\eta_p,r_{>} p)}{r_{>} p}\,,
\end{align}
where $r_<$ ($r_>$) correspond to the lesser (greater) of the coordinates $r$, $r'$, and 
%------------ Equation
\begin{align}
F_l(\eta_p,\rho)&=C_l(\eta_p)2^{-l-1}(-i)^{l+1}M_{i\eta_p, l+1/2}(2i\rho)\,,
\nonumber\\
H_l^{(+)}(\eta_p,\rho)&= (-i)^l e^{\pi\eta_p/2} 
e^{i\sigma_l(\eta_p)}
W_{-i\eta_p,l+1/2}(-i2\rho)\,,\nonumber \\
C_l(\eta_p)&=\frac{2^l e^{-\pi\eta_p/2}|\Gamma(l+1+i\eta_p)|}
{\Gamma(2l+2)}\,,
\end{align}
with conventionally defined  Whittaker functions  
$M_{k,\mu}(z)$ and $W_{k,\mu}(z)$. 
$F_l(\eta_p,\rho)$ is the regular Coulomb wave function, the irregular 
wave function is given by 
$G_l(\eta_p,\rho) = H_l^{(+)}(\eta_p,\rho) - i F_l(\eta_p,\rho)$, and 
$\sigma_{l}=\operatorname{arg}\Gamma(l+1+i\eta_p)$ is the Coulomb phase shift. 
We define the Coulomb Green's function for a bound state with binding 
energy $B$ as 
%-------------  Equation
\begin{align}
G^{(l)}_C(-B, r', r)=-i\frac{\mu\gamma}{2\pi}
\frac{F_l(\eta_{i\gamma}, i\gamma r' )}{i\gamma r'}
\frac{H^{(+)}_l(\eta_{i\gamma}, i\gamma r)}{i\gamma r}\,,
\end{align}
where $\gamma=\sqrt{2\mu B}$ is the binding momentum. 
The coordinate space definitions assume $r'<r$. In the limit 
$r'\sim 0\ll r\sim \infty$ for charged neutral particles 
($Z_\psi=0=Z_\phi$), we recover the expected result 
%------------- Equation
\begin{align}
G^{(0)}_C(-B,r'\to0,r)\sim -\frac{\mu}{2\pi r} e^{-\gamma r}\, .
\end{align}

%===========================================================
\section{Elastic scattering}\label{sec_elastic}
%===========================================================
%\input{sec4-elastic}
The elastic scattering amplitude $T(E;\bm{p}',\bm{p})$ in the presence of 
both short-range strong and long-range Coulomb interactions is traditionally 
written as 
%----------- Equations
\begin{align}
T(E;\bm{p}',\bm{p}) = T_C(E;\bm{p}',\bm{p})+T_{SC}(E;\bm{p}',\bm{p})\,,
\end{align}
where the purely Coulomb contribution can be written as 
%------------- Equation
\begin{align}
T_C(E;\bm{p}',\bm{p})&=\sum_{l=0}^\infty(2l+1) T_C^{(l)}(E;p) 
P_l(\hat{\bm{p}}'\cdot\hat{\bm{p}})
=-\frac{2\pi}{\mu}
\sum_{l=0}^\infty(2l+1) \frac{e^{2i\sigma_l}-1}{2ip} 
P_l(\hat{\bm{p}}'\cdot\hat{\bm{p}})\,,
\end{align}
using the incoming (outgoing) c.m. momentum $\bm{p}$ ($\bm{p}'$). 

The on-shell 
Coulomb-subtracted amplitude can also be expanded in partial waves as 
%------------- Equation
\begin{align}\label{eq:Tl_delta}
T_{SC}^{(l)}&=-\frac{2\pi}{\mu}\frac{e^{2i\sigma_l}}{p\cot\delta_l-ip}\,,
\end{align} 
where the full phase shift is simply $\delta_l+\sigma_l$. 
The Coulomb-subtracted phase shift $\delta_l$ is usually expressed in terms of a 
modified effective range expansion (ERE)
%------------- Equation
\begin{eqnarray}\label{eq:ERE}
\[\frac{\Gamma(2l+2)}{2^l\Gamma(l+1)}\]^2 [C_l(\eta_p)]^2 p^{2l+1}
(\cot\delta_l-i)&=&-\frac{1}{a_l}+\frac{1}{2} r_l p^2
+\frac{1}{4} s_l p^4+\cdots
\nonumber\\&&
-\frac{2k_C\, p^{2l}}{\Gamma(l+1)^2}
\frac{|\Gamma(l+1+i\eta_p)|^2}{|\Gamma(1+i\eta_p)|^2}H(\eta_p)\,,
\nonumber\\
H(\eta)&=&\psi(i\eta)+\frac{1}{2i\eta}-\ln(i\eta)\,,
\end{eqnarray}
with $\psi(x)$ the digamma function. The $\cdots$ above represents terms with higher powers in $p^2$.

%%%------------- Figure  2 --------------------------------
\begin{figure}[tbh]
\begin{center}
\includegraphics[width=0.68\textwidth,clip=true]{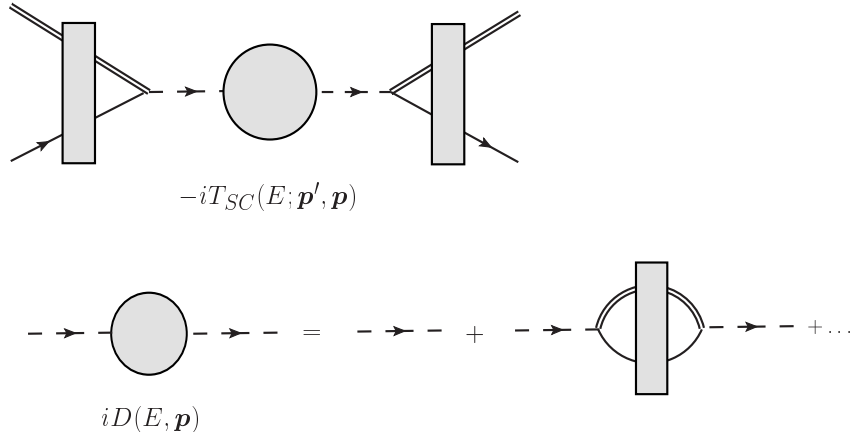} 
\end{center}
\caption{\protect Elastic scattering in $s$- and $p$-wave. 
The dashed line represents the auxiliary field $\chi$ in $s$- and $p$-wave 
as appropriate. The rest of the notation is the same as in 
Fig.~\ref{fig:CoulombLadder}.}
\label{fig:ElasticScattering}
\end{figure}
The amplitude $T_{SC}$ is given in EFT by the set of diagrams in 
Fig.~\ref{fig:ElasticScattering}. The $s$-wave amplitude for c.m. incoming 
momentum $\bm{p}$ and outgoing momentum $\bm{p}'$, with $p=|\bm p|=|\bm{p}'|$ 
and $E=p^2/(2\mu)$, can be written as
%------------- Equation
\begin{align}\label{eq:T0_EFT}
-iT_{SC}^{(0)}&=(2\mu)^2
\[\int\frac{d^3 k}{(2\pi)^3}\frac{t_c(E;\bm{p}',\bm{k})}{k^2-p^2-i\epsilon}\]
\[-i(h^{(s)})^2 D^{(s)}\left(E, 0\right)\]
\[\int\frac{d^3 l}{(2\pi)^3}\frac{t_c(E;\bm{l},\bm{p})}{l^2-p^2-i\epsilon}\]
\nonumber\\
&=-i[h^{(s)}]^2D^{(s)}(E,0)
\tilde\chi_{p'}^{(-)*}(0)\tilde\chi_{p}^{(+)}(0)
=-i[h^{(s)}]^2D^{(s)}(E,0) C_0^2(\eta_p)e^{i2\sigma_0}\,,
\end{align}
where the dressed dimer propagator is given by 
%------------- Equation
\begin{align}
D^{(s)}(p_0, \bm{p}=0)&=\frac{1}{\Delta^{(s)}+p_0-[h^{(s)}]^2 
J_0(\sqrt{2\mu p_0})}\,,\nonumber\\
J_0(p=\sqrt{2\mu E})&=G_C(E;r=0,r'=0)
=-2\mu\int\frac{d^3 q}{(2\pi)^3}
\frac{1}{\left[q^2-p^2-i\epsilon\right]}
\frac{2\pi\eta_q}{\left[e^{2\pi\eta_q}-1\right]}\,.
\end{align}

Using the modified ERE from Eq.~(\ref{eq:ERE}) in Eq.~(\ref{eq:Tl_delta}) and comparing to
the $s$-wave EFT 
expression in Eq.~(\ref{eq:T0_EFT}), we can determine the two couplings in terms of $a_0$ and $r_0$, 
%------------- Equation
\begin{align}
\Delta^{(s)}&=\frac{\mu h^{(s)2}}{2\pi a_{0}}
-\frac{\mu h^{(s)2}}{2\pi}\Bigg\{2k_C\left[\frac{1}{D\!-\!4}
-\ln\left(\frac{\lambda\sqrt{\pi}}{2k_C}\right)
-1+\frac{3}{2}C_E\right]+\lambda\Bigg\}\,,
\\
{[}h^{(s)}{]}^2&= -\frac{2\pi}{\mu^2r_{0}}\,,
\nonumber
\end{align}
where the space-time dimensions $D\to 4$ and $\lambda$ is the renormalization scale within the 
power-divergence subtraction (PDS) scheme~\cite{Kaplan:1998tg}. 
The $\lambda$-dependent 
EFT couplings were derived to reproduce 
the scattering amplitude written only in terms of scattering parameters $a_0$, $r_0$, etc. Thus the physical
observables are finite and explicitly independent of $\lambda$.

The elastic scattering amplitude requires non-perturbative treatment of the 
Coulomb photons at low energies. However, it is not clear if the short-range 
interaction contained in the parameters $a_0$, $r_0$, etc., should be 
included in perturbation or not. Fitting the EFT expression to both elastic 
and capture data 
{\em a posteriori}, 
we find that both the scattering length $a_0$ and 
the effective range $r_0$ contribute at LO. 
We propose a power-counting where $a_0\sim \Lambda^2/Q^3$ is fine-tuned, and 
$r_0\sim 1/\Lambda$ is of natural size. However, there is a further fine-tuning in that the combination 
$r_0 p^2/2-2k_C H(\eta_p)\sim Q^3/\Lambda^2$ for low momentum 
$p\sim Q$~\cite{Higa:2008dn}. 
Then the $s$-wave scattering amplitude gets LO contributions from $a_0$, 
$r_0$ for $p\sim Q$. The contribution from a natural-size 
shape parameter $s_0\sim 1/\Lambda^3$ is suppressed by a relative factor of 
$Q/\Lambda$, and contributes at NLO.

The $p$-wave amplitude is written as 
%------------- Equation
\begin{align}\label{eq:T1_EFT}
-i 3 T_{SC}^{(1)} \hat{\bm p}'\cdot\hat{\bm p}=-i[h^{(\zeta)}]^2D^{(\zeta)}(E,0)
[\nabla_a{\tilde\chi_{p'}^{(-)\ast}}(0)][\nabla_a{\tilde\chi_p^{(+)}(0)}]\,,
\end{align}
with the $p$-wave dressed dimer propagator written as
%------------- Equation
\begin{align}\label{eq:dimerPwave}
D^{(\zeta)}(E,0)=&\frac{1}{\Delta^{(\zeta)}
+E-3[h^{(\zeta)}]^2 J_1(p)/\mu^2}\,,\nonumber\\
J_1(p)=&-\frac{2\mu}{n}\int \frac{d^n k}{(2\pi)^n}\frac{1}{k^2-p^2-i\delta}
[\nabla_a\tilde\chi_k^{(+)\ast}(0)][\nabla_a\tilde\chi^{(+)}_k(0)]\,.
\end{align}
$J_1(p)$ is given by a divergent integral that we regulate through 
dimensional regularization in 
$n=D-1$ space dimensions. 
Using the relations 
%------------- Equation
\begin{align}
[\nabla_a{\tilde\chi_{p'}^{(-)\ast}} (0)][\nabla_a{\tilde\chi_p^{(+)}(0)}]=& 
e^{-\pi\eta_p}\Gamma(2+i\eta_p)^2 \,\bm{p}'\cdot\bm{p}\,,\nonumber\\
[\nabla_a\tilde\chi_k^{(+)\ast}(0)][\nabla_a\tilde\chi^{(+)}_k(0)]
=&(k^2+\beta^2)\frac{2\pi\eta_k}{e^{2\pi\eta_k}-1}\,,
\end{align} 
the PDS prescription for the integrals, and relating Eqs.~(\ref{eq:Tl_delta}), (\ref{eq:ERE}) 
to Eq.~(\ref{eq:T1_EFT}), one gets 
%------------- Equation
\begin{eqnarray}\label{eq:pwavecouplings}
\Delta^{(\zeta)}&=&\frac{h^{(\zeta)2}}{2\pi\mu a_{1}}
-\frac{h^{(\zeta)2}}{2\pi\mu}\Bigg\{2k_C^3\left[\frac{1}{D\!-\!4}
-\ln\left(\frac{\lambda\sqrt{\pi}}{2k_C}\right)
-1+\frac{3}{2}C_E\right]
\nonumber\\&&
+k_C^2\left(\frac{3\lambda}{2}-\frac{2k_C}{3}\right)
+8\pi^2k_C^3\zeta'(-2)+\frac{\pi^2\lambda k_C^2}{2}-\frac{3\pi\lambda^2 k_C}{2}
+\frac{\pi\lambda^3}{2}\Bigg\}\,,\nonumber\\
\frac{2\pi}{[h^{(\zeta)}]^2}&=& -r_1^{(\zeta)}
-2\Bigg\{2k_C\left[\frac{1}{D\!-\!4}
-\ln\left(\frac{\lambda\sqrt{\pi}}{2k_C}\right)
-1+\frac{3}{2}C_E\right]
+\left(\frac{3\lambda}{2}-\frac{2k_C}{3}\right)\Bigg\}\,.
\end{eqnarray}
The $p$-wave amplitude is $\lambda$-independent though the EFT couplings evolve as functions of it, similar to the $s$-wave result above.

The dressed $p$-wave dimer propagator defines the wave function 
renormalization constant 
%------------- Equation
\begin{align}
\frac{1}{{\cal Z}^{(\zeta)}}=&\frac{\partial}{\partial p_0}
\left[D^{(\zeta)}(p_0;{\mbox{\boldmath $p$}})^{-1}
\right]\big|_{p_0=p^2/(2\mu)-B^{(\zeta)}}
\nonumber\\=&
-\frac{{h^{(\zeta)}}^2}{2\pi p}\frac{\partial}{\partial p}
\left[ 
-\frac{1}{a_1}+\frac{1}{2} r_1 p^2 +\frac{1}{4} s_1 +\cdots
-2k_C(k_C^2+p^2) H(\eta_p)
\right]\big|_{p=i\gamma}\,,
\label{eq:wfrenormdef}
\end{align}
where 
$B^{(\zeta)}$ is the $p$-wave binding energy.  To ensure the modified ERE 
parameters are consistent with the bound state energies
we redefine the ERE for $p$-waves as 
%------------- Equation
\begin{eqnarray}
9 [C_1(\eta_p)]^2 p^3(\cot\delta_1-i)&=&2 k_C(k_C^2-\gamma^2)H(-i\eta_\gamma)
+\frac{1}{2}\rho_1(p^2\!+\!\gamma^2)\!+\!
\frac{1}{4}\sigma_1(p^2\!+\!\gamma^2)^2
+\cdots \nonumber\\&&
-2 k_C(k_C^2+p^2) H(\eta_p)\,.
\end{eqnarray}
A straightforward calculation leads to 
%------------- Equation
\begin{align}
-\frac{2\pi}{h^{(\zeta)2}{\cal Z}^{(\zeta)}}&
=\rho_1^{(\zeta)}-4k_C\,H\left(-i\frac{k_C}{\gamma}\right)
-\frac{2k_C^2}{\gamma^3}(k_C^2-\gamma^2)
\left[\psi'\left(\frac{k_C}{\gamma}\right)
-\frac{\gamma^2}{2k_C^2}-\frac{\gamma}{k_C}\right]\,,
\label{eq:wfrenormdef2}
\end{align}
such that the wave function renormalization constant  depends only on the 
binding momentum $\gamma$ and the effective range $\rho_1$, to all orders 
in perturbation.  From the second relation in Eq.~(\ref{eq:pwavecouplings}), 
we propose $\frac{2\pi}{[h^{(\zeta)}]^2}\sim Q\sim k_C\sim r_1^{(\zeta)}\sim 
\rho_1^{(\zeta)}$. 
Then the first relation gives $\Delta^{(\zeta)}\sim Q^2/\mu$, and all the 
terms in the denominator of the dimer propagator  in Eq.~(\ref{eq:dimerPwave}) 
scale as $Q^2/\mu$. This makes the contribution from the binding momentum 
$\gamma$ and effective range $\rho_1$ LO. 
The combination  $h^{(\zeta)2}{\cal Z}^{(\zeta)}$ controls the 
normalization of the capture cross-section as shown in the next section. 
The shape parameter $\sigma_1$ does not contribute to the wave function 
renormalization or the capture cross section. However, a natural-size 
$\sigma_1 \sim1/\Lambda$ would contribute to the $p$-wave phase shift at NLO.

%===========================================================
\section{Radiative capture}\label{sec_capture}
%===========================================================
%\input{sec5-capture}
We assign the c.m. momenta $\bm{p}$ to the $\alpha$ particle, and $\bm{k}$ 
to the outgoing photon in the final state. From energy-momentum conservation 
$|\bm{k}|=(p^2+\gamma^2)/(2\mu)\sim Q^2/\mu\ll Q\sim\gamma\sim p$ in the EFT 
power-counting. Thus in a typical loop calculation a combination such as 
$E_{\psi}+k_0 + q_0 -(\bm{q}+\bm{k}+\bm{p})^2/(2m_\psi)$ is approximated as 
$E_{\psi} + k_0 +q_0 -(\bm{q}+\bm{p})^2/(2m_\psi)\sim Q^2/\mu$ where 
$E_{\psi}=p^2/(2m_\psi)$ and $(q_0,\bm{q})$ is the loop energy-momentum. 
This approximation corresponds to zero-recoil of the final bound state $^7$Be. 
We count $Q/\mu\sim Q^2/\Lambda^2\ll Q/\Lambda$ and neglect recoil effects in this calculation 
up to NLO. 

\begin{figure}[tbh]
\begin{center}
\includegraphics[width=0.58\textwidth,clip=true]{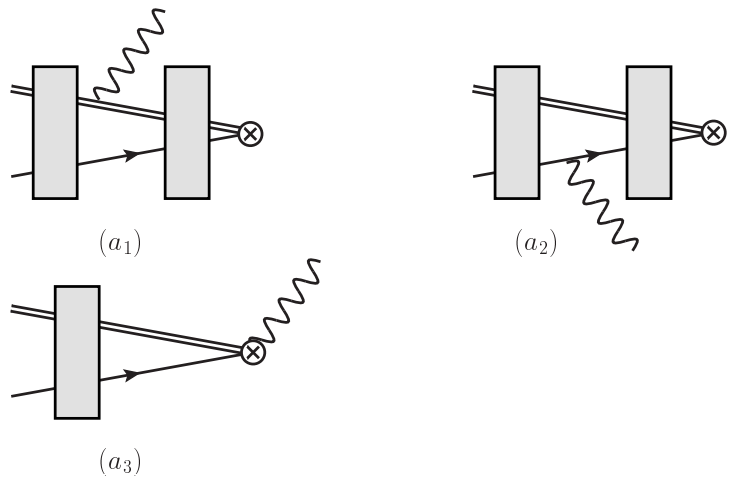} 
\end{center}
\caption{\protect Group A diagrams. Radiative capture without initial state 
strong interaction. The $\otimes$ represents the final bound state. The 
external photon represented by the wavy line is minimally coupled to the 
charged particles as described in the text. The rest of the notation is 
the same as in Figs~\ref{fig:CoulombLadder}, ~\ref{fig:ElasticScattering}.}
\label{fig:CaptureTriangle}
\end{figure}

\begin{figure}[tbh]
\begin{center}
\includegraphics[width=0.69\textwidth,clip=true]{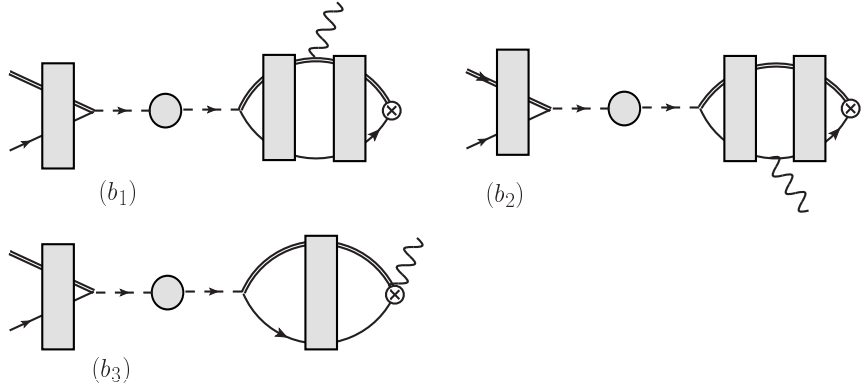} 
\end{center}
\caption{\protect Group B diagrams. Radiative capture with initial state 
$s$-wave strong interaction represented by the dressed dimer. We use the same 
notation as Figs.~\ref{fig:CoulombLadder}, ~\ref{fig:ElasticScattering} 
and ~\ref{fig:CaptureTriangle}.}
\label{fig:CaptureBubble}
\end{figure}

\begin{figure}[tbh]
\begin{center}
\includegraphics[width=0.47\textwidth,clip=true]{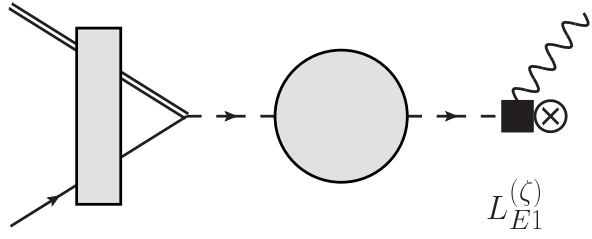} 
\end{center}
\caption{\protect Two-body current represented by the square. The rest of the 
notation is the same as in Figs~\ref{fig:CoulombLadder},
~\ref{fig:ElasticScattering},~\ref{fig:CaptureTriangle}.}
\label{fig:TwoBodyCurrent}
\end{figure}

In this section we present some of the Feynman diagrams that contribute 
to the capture process, see Figs.~\ref{fig:CaptureTriangle},
~\ref{fig:CaptureBubble}, ~\ref{fig:TwoBodyCurrent}.
In the next section when we present our analysis, we elaborate more on the 
power-counting and discuss how these sets of diagrams constitute the EFT 
contribution up to NLO. We consider E1 transitions from the initial 
$^2S_{1/2}$ state to both final bound states, ground $^2P_{3/2}$ and 
excited $^2P_{1/2}$. The external photon is minimally coupled to the charged 
clusters at LO. Capture from initial $^2D_{5/2}$ and $^2D_{3/2}$ states without strong interaction are included as well.

%%%------------- Figure 3 --------------------------------
The first set of diagrams  from Fig.~\ref{fig:CaptureTriangle} include only Coulomb interactions for the incoming 
charged particles $^3$He and $\alpha$. We find for $s$-wave capture
%----------- Equation
\begin{align}\label{eq:diagramA}
(a_1)+(a_2)&=
-\frac{e^{i\sigma_0}\,A(p)}{\mu}C_0(\eta_p)\Gamma_{aa}^{(\zeta)}\,,
\nonumber\\
(a_3)&= 
-\tilde\chi_p^{(+)}({\bm r}=\bm{0}) \Gamma_{aa}^{(\zeta)}
=-C_0(\eta_p)e^{i\sigma_0}\,\Gamma_{aa}^{(\zeta)}\,,\nonumber\\
A(p)&= 
\frac{2\gamma\mu}{3}\frac{\Gamma(2+k_C/\gamma)}{C_0(\eta_p)}
\int_0^\infty dr\,r 
W_{-k_C/\gamma,3/2}(2\gamma r)\partial_r
\left[\frac{F_0(k_C/p)}{p r}\right]\,,
\end{align}
where the index $\zeta$ refers to the  final 
%state 
$p$-wave bound states. 
The corresponding binding momentum $\gamma$ is given by 
$\gamma_0=\sqrt{2\mu B_0}$ and $\gamma_1=\sqrt{2\mu B_1}$ for the ground and 
the first excited states, respectively. The Whittaker function 
$W_{-k_C/\gamma,3/2}(2\gamma r)$ is associated with the final $p$-wave 
bound state, and the $s$-wave Coulomb wave function $F_0(\eta_p)$ is 
associated with the initial incoming scattering state.  There is also a $d$-wave contribution from diagrams $(a_1)+(a_2)$ that is given by
%-----------  Equation
\begin{align}\label{eq:dwaveAmplitude}
&-3\mu e^{i\sigma_2}Y(p)(\hat{p}_a\hat{p}_b-\frac{1}{3}\delta_{ab})\Gamma_{ab}\,, \ \ \mathrm{with}
\nonumber\\
&Y(p)=  \frac{2\gamma}{3}\Gamma(2+k_C/\gamma)\int_0^\infty dr\  r W_{-k_C/\gamma,3/2}(2\gamma r)
\(\frac{\partial}{\partial r} +\frac{3}{r}\) \frac{F_2(\eta_p, r p)}{r p}\,,
\end{align}
where we chose the incoming relative momentum $\bm{p}$ to point in the $\hat{z}$-direction. 

The projection onto the $p$-wave states is given by 
%-----------  Equation
\begin{align}
\Gamma^{(\zeta)}_{ab}=\(\frac{e Z_\phi}{m_\phi}
-\frac{e Z_\psi}{m_\psi}\)(h^{(\zeta)}\sqrt{3}\sqrt{\mathcal Z^{(\zeta)}}
\sqrt{2m_\phi})
\epsilon_a^\ast {U_{i}^{\ast\alpha,\zeta}}
(-\vec{\bm k}) P_{ib}^{\alpha\beta,\zeta}U^{\beta,\psi}(-\vec{\bm p})\,,
\end{align}
where $\epsilon_a$ is the photon polarization vector, 
$\mathcal Z^{(\zeta)}$ is given by Eq.~(\ref{eq:wfrenormdef2}), 
$U^\zeta$ is the spinor field for the $p$-wave final state with 
mass $M=m_\psi+m_\phi$, and $U^\psi$ is the spinor field for the incoming 
$^3$He nucleus. The spinor fields satisfy the completeness relations 
%------------ Equation
\begin{align}
  \sum_\mathrm{pol.} U_i^{\alpha,\zeta}(\bm{p})[U_j^{\beta,\zeta}(\bm{p})]^\ast
  &=2 M P_{ij}^{\alpha\beta,\zeta}\,,\nonumber\\
 \sum_\mathrm{pol.} U^{\alpha,\psi}(\bm{p})[U^{\beta,\psi}(\bm{p})]^\ast
  &=2 m_\psi\delta^{\alpha\beta}\,,
\end{align}
where $i$, $j$ are vector indices, and $\alpha$, $\beta$ are spin indices. 

%%%------------- Figure 4 --------------------------------
The second set of diagrams from Fig.~\ref{fig:CaptureBubble} involve the 
initial state short-range interaction that is constrained by the $s$-wave 
phase shift through the ERE. We find
%----------- Equation
\begin{align}
(b_1)+(b_2)&=
\frac{2\pi}{\mu}\frac{C_0(\eta_p)e^{i\sigma_0} }
{-\frac{1}{a_0}+\frac{r_0}{2} p^2+\frac{s_0}{4} p^4+\cdots -2k_C H(\eta_p)}
\frac{B_{ab}(p)}{\mu}\Gamma_{ab}\,,\nonumber\\
B_{ab}(p)&=-3\int d^3 r
\[\frac{G_C^{(1)}(-B;r',r)}{r'}\]\Big|_{r'=0}
\frac{\partial G_C^{(+)}(E;\bm{r},0)}{\partial r_a}\frac{r_b}{r}\,,
\nonumber\\
\[\frac{G_C^{(1)}(-B;r',r)}{r'}\]\Big|_{r'=0}&=-\frac{\mu\gamma}{6\pi r}
\Gamma(2+k_C/\gamma)
W_{-k_C/\gamma,3/2}(2\gamma r)\,,
\nonumber\\
G_C^{(+)}(E;\bm{r},0)&=-\frac{\mu}{2\pi r}\Gamma(1+ik_C/p)
W_{-ik_C/p,1/2}(-i2 p r)\,.
\end{align}
The integral $B_{ab}$ is divergent, which is rendered finite when combined 
with the contribution from the third diagram, 
%-----------  Equation
\begin{align}
(b_3)&= \frac{2\pi}{\mu}\frac{C_0(\eta_p)e^{i\sigma_0} }
{-\frac{1}{a_0}+\frac{r_0}{2}p^2+\frac{s_0}{4} p^4+\cdots -2k_C H(\eta_p)}
J_0(p)\Gamma_{aa}\,.
\end{align}
We regulate the divergences using PDS, which is most conveniently done in 
this calculation in momentum space. The divergences come from zero and 
single Coulomb photon exchanges. Thus we analytically calculate the 
divergent pieces perturbatively up to ${\cal O}(\alpha_e)$ and calculate the 
rest (more than a single Coulomb photon), that is not divergent, numerically: 
%-----------  Equation
\begin{align}\label{eq:Bpert}
B_{ab}(p)&\equiv B_{ab}^{(0)}(p)+\alpha_e B_{ab}^{(1)}(p)+\Delta B_{ab}(p)\,,
\nonumber\\
B_{ab}^{(0)}&\equiv B^{(0)}\delta_{ab}=\mu^2\left[
\frac{\lambda}{2\pi}
+\frac{1}{3\pi}\frac{i p^3-\gamma^3}{p^2+\gamma^2}
\right]\delta_{ab}\,,\nonumber\\
\alpha_e B_{ab}^{(1)}&\equiv \alpha_e B^{(1)}\delta_{ab}=
-\frac{k_C\mu^2}{2\pi}\left[
\frac{1}{4-D}+\ln\frac{\pi\lambda^2}{4k_C^2}
-\gamma_E-\frac{2}{3}+\ln4\pi
\right]\delta_{ab}+ k_C\, C(p)\delta_{ab}\,,\nonumber\\
C(p)&=\frac{\mu^2}{6\pi^2(p^2+\gamma^2)}\int_0^1dx\int_0^1dy
\frac{1}{\sqrt{x(1-x)}\sqrt{1-y}}\nonumber\\
&\times \( xp^2\ln\left[\frac{\pi}{4k_C^2}
(-yp^2+(1-y)\gamma^2/x-i\delta) \right] \right.\nonumber\\
&\left. +p^2 \ln\left[\frac{\pi}{4k_C^2}
(-yp^2-(1-y)p^2/x-i\delta) \right]  \right.\nonumber\\
&\left. +x\gamma^2\ln\left[\frac{\pi}{4k_C^2}
(y\gamma^2+(1-y)\gamma^2/x-i\delta) \right]  \right.\nonumber\\
&\left. +\gamma^2\ln\left[\frac{\pi}{4k_C^2}(y\gamma^2-(1-y)
p^2/x-i\delta) \right] 
\right)\,.
\end{align}
The double integral $C(p)$ can be reduced further to a single integral that 
we evaluate numerically. The finite piece $\Delta B_{ab}$ is evaluated 
numerically where we use spherical symmetry to write 
$(r_b/r)[\partial/\partial r_a]=(r_ar_b/r^2)[\partial/\partial r]
\rightarrow (\delta_{ab}/3)[\partial/\partial r]$ in the integral. 
Consequently, $\Delta B_{ab}(p)\equiv \Delta B(p)\delta_{ab}$ and also $B_{ab}(p)\equiv B(p) \delta_{ab}$. 

We do a similar decomposition of $J_0(p)$ to write
%-----------  Equation
\begin{align}\label{eq:J0pert}
J_0(p)&= J_0^{(0)}(p)+\alpha_e J_0^{(1)} +\Delta J_0(p)\,,\nonumber\\
J_0^{(0)}(p)&=-\frac{\mu}{2\pi}(ip+\lambda)\,,\nonumber\\
\alpha_e J_0^{(1)}&=\frac{k_C\mu}{2\pi}\left[\frac{1}{4-D}
+1-\gamma_E+i\pi+\ln\frac{\pi\lambda^2}{4 p^2}\right]\,,
\end{align}
and the finite piece as
%-----------  Equation
\begin{align}
\Delta J_0(0)&= -2\mu \int\frac{d^n q}{(2\pi)^n}\frac{1}{q^2-p^2-i\delta}
\left[ \frac{2\pi\eta_q}{\exp(2\pi\eta_q)-1} -1
+\pi\eta_q\right]\nonumber\\
&=\frac{i\mu p}{2\pi}+\frac{k_C\mu}{2\pi}
\left[-2 H(\eta_p) -2\gamma_E -i\pi
+\ln\frac{p^2}{k_C^2} \right]\,.
\end{align}

Comparing Eqs.~(\ref{eq:Bpert}) and (\ref{eq:J0pert}), we see that the 
divergent pieces in the term $J_0(p)+ B(p)/\mu$  cancel when combining 
diagrams $(b_1)+(b_2)$ with $(b_3)$. The final result is finite and 
independent of the renormalization scale $\lambda$.

%%%------------- Figure 5 --------------------------------
The third contribution is from two-body currents in 
Eq.~(\ref{eq:TwoBodyCurrentL1}) that contribute to E1 
capture to the ground and excited states. The diagram in 
Fig.~\ref{fig:TwoBodyCurrent} gives the contribution
%-----------  Equation
\begin{align}\label{eq:TwoBodyCurrent}
\frac{2\pi C_0(\eta_p)e^{i\sigma_0}\; k_0}{-1/a_0 +\frac{r_0}{2} p^2
+\frac{s_0}{4} p^4+\cdots -2k_C H(\eta_p)}\times
L^{(\zeta)}_{E1}\,.
\end{align}
Thus we have the following expression for the capture amplitude from the initial $s$-wave state
%-----------  Equation
\begin{eqnarray}\label{eq:S34amplitude}
\mathcal{A}(p)\Gamma^{(\zeta)}_{aa}
&\equiv&
-e^{i\sigma_0} C_0(\eta_p)\,
\Gamma^{(\zeta)}_{aa}\left[\frac{1}{\mu}A(p)+1
-\frac{2\pi}{\mu^2}
\frac{B^{(0)}(p) +\alpha_e B^{(1)}(p)+\Delta B(p)+ \mu J_0(p) }
{-\frac{1}{a_0}+\frac{r_0}{2} p^2+\frac{s_0}{4} p^4+\cdots-2k_C H(\eta_p)}
\right.\nonumber\\
&&\left.
-\frac{2\pi k_0}
{-1/a_0+\frac{r_0}{2} p^2+\frac{s_0}{4}p^4+\cdots-2k_C H(\eta_p)}\times
L^{(\zeta)}_{E1}
\right]\,.
\end{eqnarray}
  
The total  cross section and the $S$-factor can be calculated as 
%-----------  Equation
\begin{align}\label{eq:Sfactor}
\sigma(p)&=\frac{1}{16\pi M^2}\frac{1}{2}
\[ \frac{p^2+\gamma_1^2}{2\mu p} |M^{(^2P_{1/2})}|^2
+\frac{p^2+\gamma_0^2}{2\mu p}|M^{(^2P_{3/2})}|^2
\]\,,\nonumber\\
S_{34}(E)&= E e^{2\pi\eta_p}\sigma(p=\sqrt{2\mu E})\,,
\end{align}
where we averaged over the initial $^3$He spin and summed over the final state photon polarization and $^7$Be spin. 
The amplitude squared is
%-----------  Equation
\begin{eqnarray}
|M^{(\zeta)}|^2&=&(2j+1)
\(\frac{ Z_\phi m_\psi}{M}-\frac{ Z_\psi m_\phi}{M}\)^2
\frac{64\pi\alpha_e  M^2 ([h^{(\zeta)}]^2 \mathcal Z^{(\zeta)})}{\mu}
\nonumber\\
&&\times \[ |\mathcal A(p)|^2 +2|Y(p)|^2\]\,,
\end{eqnarray}
with angular momentum $j=1/2$ for the excited state and $j=3/2$ for the 
ground state.  The function $Y(p)$ is the contribution from $d$-wave initial states, Eq~(\ref{eq:dwaveAmplitude}). These expressions reduce to the corresponding forms for the 
capture reaction \nLi ~when the long-range Coulomb interaction is turned 
off~\cite{Rupak:2011nk,Fernando:2011ts}.

%===========================================================
\section{Results and Analysis}\label{sec_analysis}
%===========================================================
%\input{sec6-analysis}
To predict the $S$-factor at solar energies using the EFT expression in 
Eq.~(\ref{eq:Sfactor}) we need to determine the elastic $s$- and $p$-wave 
scattering parameters, and the  two couplings for two-body currents. Usually, 
one-body currents dominate. Thus we start the analysis with the diagrams 
from Figs.~\ref{fig:CaptureTriangle} and ~\ref{fig:CaptureBubble} that can 
be constrained by elastic scattering, in principle. 
However, the phase shifts for $^3$He$(\alpha,\alpha)^3$He
are poorly known. 
The phase shift analysis is from a very old 
source~\cite{Spiger:1967} that was 
analyzed again by Boykin {\em et al.} in Ref.~\cite{Boykin:1972} where the experimental errors were 
quantified.

Numerical analysis of the $s$-wave phase shifts yields $a_0\sim 20-30$ fm and $r_0\sim 1$ fm. 
The size of the effective range $r_0$ is consistent with $\Lambda\sim 1/r_0\sim 150 -200$ MeV which is about the pion mass, the expected breakdown scale of the halo EFT. 
We take the low-momentum scale $Q$ to be about the binding momenta of the 
ground and excited states, $\gamma_0\sim \gamma_1\sim 60- 70$ MeV. 
Thus this suggests the scattering length to be 
fine-tuned. 
At an arbitrarily low momentum, the relative contribution of the diagrams from 
Fig.~\ref{fig:CaptureBubble} compared to the diagrams in 
Fig.~\ref{fig:CaptureTriangle}, see Eq.~(\ref{eq:S34amplitude}), is 
$\sim 2\pi (B+\mu J_0) a_0/\mu^2$. In the absence of Coulomb interaction 
$2\pi (B+\mu J_0) /\mu^2\sim Q$. However, this naive expectation is 
invalidated due to the non-perturbative Coulomb contributions~\cite{Higa:2008dn}, and instead we 
find $2\pi (B+\mu J_0) /\mu^2\sim 1/a_0\sim Q^3/\Lambda^2$.  Therefore the contribution from the second set of diagrams scales as $\sim 2\pi (B+\mu J_0) a_0/\mu^2\sim 1$ compared to the first set. This scaling holds over a range of momenta $p\lesssim 70$ MeV. 

The relative contribution of the two body currents scales as 
$\sim a_0 k_0 L_\mathrm{E1}^{(\zeta)}$. The photon energy $k_0$ scales as 
$\sim Q^3/\Lambda^2$ since we count 
$p/\mu\sim\gamma/\mu\sim Q/\mu\sim Q^2/\Lambda^2$~\cite{Higa:2008dn}. 
The relative contribution of the two body current is 
$\sim a_0 Q^3 L_\mathrm{E1}^{(\zeta)}/\Lambda^2\sim L_\mathrm{E1}^{(\zeta)}$ 
for $a_0\sim \Lambda^2/Q^3$. For natural-size couplings 
$L_\mathrm{E1}^{(\zeta)}\sim 1$, the two body currents also contribute at LO. 

The NLO corrections to the capture amplitude come from the $s$-wave shape parameter corrections assuming $s_0\sim 1/\Lambda^3$. The amplitude from Eq.~(\ref{eq:S34amplitude}) can now be expanded as
%----------- Equation
\begin{align}
\mathcal{A}(p)=&\mathcal{A}_\mathrm{LO}(p)+\mathcal{A}_\mathrm{NLO}(p)+\dots\,,
\nonumber\\
\mathcal{A}_\mathrm{LO}(p)=&
e^{i\sigma_0} C_0(\eta_p)\left[
-\frac{A(p)}{\mu}-1+\frac{2\pi}{\mu^2}
\frac{B^{(0)}(p) +\alpha_e B^{(1)}(p)+\Delta B(p)+ \mu J_0(p) +\mu^2 k_0  L^{(\zeta)}_{E1}}
{-\frac{1}{a_0}+\frac{r_0}{2} p^2-2k_C H(\eta_p)} \right]\,,\nonumber\\
\mathcal{A}_\mathrm{NLO}(p)=&-e^{i\sigma_0} C_0(\eta_p)\,
\frac{2\pi }{\mu^2}\frac{s_0 p^4}{4}
\frac{B^{(0)}(p) +\alpha_e B^{(1)}(p)+\Delta B(p)+ \mu J_0(p) +\mu^2 k_0  L^{(\zeta)}_{E1}}
{[-\frac{1}{a_0}+\frac{r_0}{2} p^2-2k_C H(\eta_p)]^2}\,.
\end{align}

Naively, $d$-wave contributions would be NNLO, as they are suppressed by two relative powers of momentum compared to capture from $s$-waves. In our fits this is confirmed at c.m. energies $\lesssim 500$ keV ($p\lesssim 30$ MeV).  At higher energies ($p\sim 60$ MeV) $d$-wave contributions are around $\sim 25\%$. This might be related to an accidental cancellation between the $s$-wave capture diagrams from Figs. ~\ref{fig:CaptureTriangle} and ~\ref{fig:CaptureBubble} when $a_0\sim 30$ fm. We include the $d$-wave contribution in perturbation at NLO. Higher order two-body currents, and possible $d$-wave initial state strong interactions, would constitute NNLO errors.   

 We do a simultaneous fit to 
$S$-factor measurements and phase shift analysis to determine the parameters.  At LO, the scattering parameters $a_0$, $r_0$, $r_1^{(+)}$,
$r_1^{(-)}$ and two-body current couplings $L_\mathrm{E1}^{(+)}$ and $L_\mathrm{E1}^{(-)}$  describe the $s$-wave phase shift $\delta_0$, $p$-wave phase shifts
 $\delta_1^{(+)}$, $\delta_1^{(-)}$ and the capture cross sections.  At NLO, three additional scattering parameters $s_0$, $\sigma_1^{(+)}$, $\sigma_1^{(-)}$ contribute to the respective phase shifts. Only $s_0$ contributes to the capture process. 

The capture data are from LUNA~\cite{LUNA}, Seattle~\cite{Seattle}, 
Weizmann~\cite{Weizmann}, ERNA~\cite{ERNA}, and Notre Dame~\cite{NotreDame}. 
The $S_{34}$ measurements include both prompt photon and activation data. 
For the former, the branching ratio $R_0$ to the excited state compared to 
the ground state is available. The $R_0$ data are useful as they remove 
normalization errors from the cross section measurements. 
We perform two sets of fits. 

First, we fit the $S_{34}$ and $R_0$ to c.m. energies of about $500$ keV. Though it would be preferable to fit the ERE parameters at low energies, the available phase shift data starts at around 1.9 MeV. For the $p$-wave, the binding momenta $\gamma_0$ and $\gamma_1$ provide a constraint at zero energy. For the $s$-wave, no such low-energy constraints exist. It leads to large errors in fitting $a_0$. At LO, we fit the phase shifts to about 2.5 MeV. At NLO, we increase the range of $s$-wave phase shift fit to about 3 MeV as we introduce the shape parameter $s_0$.  We also use the LO value for $r_0$ in the NLO fits. When we let $r_0$ vary, we get similar values and fitting errors. However, the scattering length $a_0$ and shape parameter $s_0$ have large uncertainties though their central values are reasonable. Lack of low-energy phase shift information provides no meaningful constraints on $a_0$. 
We call this set of   fits ``Small range'' 
in the plots. Second, we fit the $S_{34}$ and $R_0$ to c.m. energies of about 
$1000$ keV, and all the elastic  phase shifts to about 3 MeV. We call this 
the ``Large range'' fit in the plots. Again we keep $r_0$ fixed to its LO value in the NLO fits. 
We also performed a Jackknife fit over the Large range, where we removed one whole data set from 
 each of the five experimental groups LUNA, Seattle, Weizmann, ERNA and Notre Dame, in turn. This gave results very similar to the Large range. 

%------------- Table 1 ----------------------
%\begin{widetext}
\begin{table}[htb]
\centering
\begin{adjustbox}{width=0.97\textwidth,center}
%\begin{ruledtabular}
\begin{tabular}{|l|c|c|c|c|c|c|c|c|c|}\hline
Fits & $a_0$ (fm) & $r_0$ (fm) & $s_0$ (fm$^3$) &  $\rho_1^{(+)}$ (MeV) & $\sigma_1^{(+)}$ (fm)& $\rho_1^{(-)}$ (MeV) & $\sigma_1^{(-)}$  (fm) & $L_1^{(+)}$  & $L_1^{(-)}$
\begin{filecontents*}{table_parameter_fitsdwave.csv}
Fits,a,da,r,dr,sz,dsz,ra,dra,sa,dsa,rb,drb,sb,dsb,La,dLa,Lb,dLb
Small range (LO),19.8,7.9,1.1,0.2,0,0,-55.4,0.8,0,0,-40.6,0.8,0,0,0.81,0.11,0.74,0.13
Large range (LO),17.6,4.2,1.2,0.1,0,0,-56.1,0.4,0,0,-42.6,0.5,0,0,0.85,0.05,0.94,0.06
Small range (NLO),24.8,4.7,1.1,0.2,-0.9,0.8,-57.8,1.7,1.68,0.07,-41.4,1.3,1.69,0.08,0.97,0.11,0.81,0.13
Large range (NLO),21.6,3.4,1.2,0.1,-0.9,0.7,-55.4,0.5,1.59,0.03,-41.9,0.7,1.74,0.05,0.79,0.06,0.83,0.08
\end{filecontents*}
\csvreader[head to column names]{table_parameter_fitsdwave.csv}{}
{ \\\hline\Fits &\a $\pm$ \da& \r $\pm$ \dr&
\ifthenelse{\equal{\sz}{\dsz}}{---}{\sz$\pm$\dsz}
&\ra$\pm$\dra & \ifthenelse{\equal{\sa}{\dsa}}{---}{\sa$\pm$\dsa} &\rb$\pm$\drb &
\ifthenelse{\equal{\sb}{\dsb}}{---}{\sb$\pm$\dsb}
 & \La$\pm$\dLa  &\Lb$\pm$\dLb }\\\hline
\end{tabular}
%\end{ruledtabular}
\end{adjustbox}
\caption{EFT parameters. The parameters were determined from simultaneous 
fits to the capture \HeAlpha, ~and $^3$He$(\alpha,\alpha)^3$He 
phase shift data as described in the text.  We use the LO $r_0$ values in the NLO fits.}
 \label{table:fits}
\end{table}  
%\end{widetext}

%%%------------- Figure 6 --------------------------------
\begin{figure}[tbh]
\begin{center}
  \includegraphics[width=0.69\textwidth,clip=true]{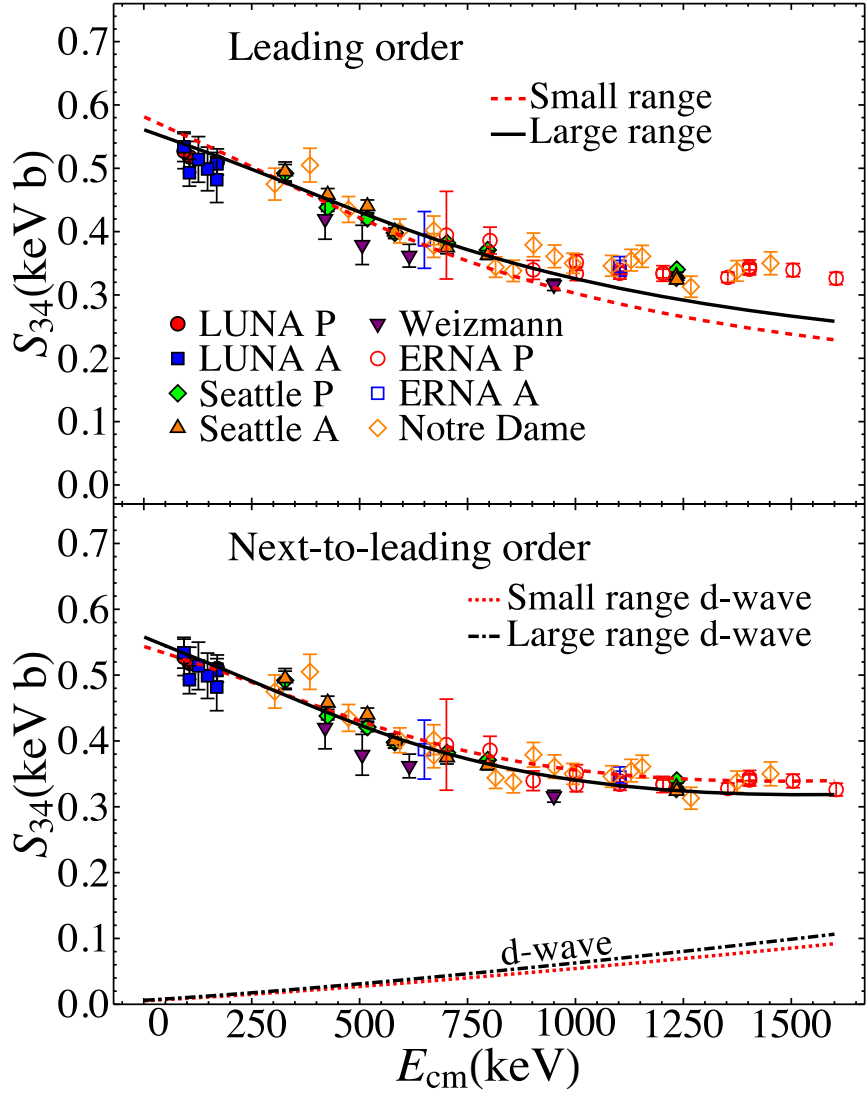}\end{center}
\caption{\protect $S$-factor $S_{34}$ for c.m. energy $E_\mathrm{cm}$~at LO (top panel) and 
NLO (bottom panel). 
The capture data are from LUNA~\cite{LUNA}, Seattle~\cite{Seattle}, 
Weizmann~\cite{Weizmann}, ERNA~\cite{ERNA}, and Notre Dame~\cite{NotreDame}. 
The letters P and A after the experiment name are used to distinguish between prompt and activation data as appropriate. The 
red dashed, and black solid curves are the halo EFT results 
from two different fits described in the text. We show the $d$-wave contributions at NLO as red dot and black dot-dashed  curves in the bottom panel for the two different fits. 
}
\label{fig:CaptureFits}
\end{figure}

%%%------------- Figure 7 --------------------------------
\begin{figure}[tbh]
\begin{center}
  \includegraphics[width=0.69\textwidth,clip=true]{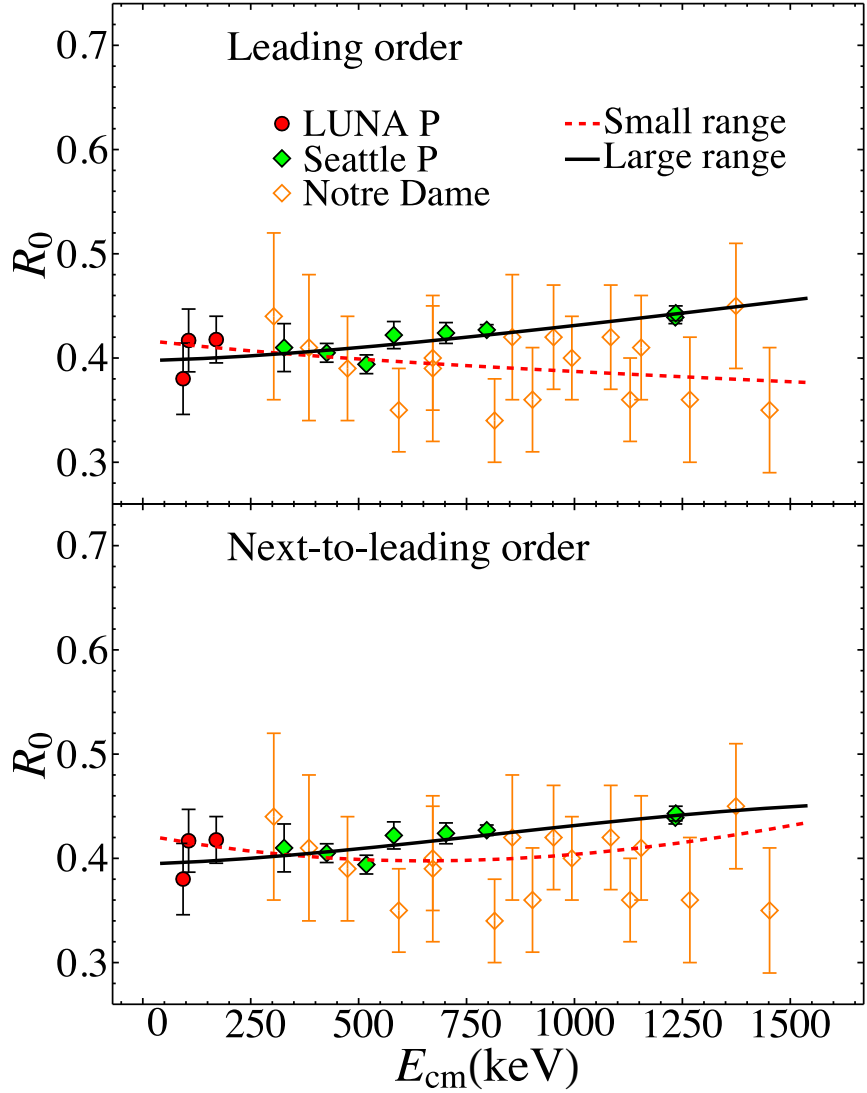}\end{center}
\caption{Branching ratio $R_0$ for capture to excited state to ground state  for \HeAlpha
~at LO (top panel) and  NLO (bottom panel). 
Data are from LUNA~\cite{LUNA}, 
Seattle~\cite{Seattle}, and Notre Dame~\cite{NotreDame}. 
The rest of the notation is the same as 
Fig.~\ref{fig:CaptureFits}.}
\label{fig:RatioFits}
\end{figure}

%%%------------- Figure 8 --------------------------------
\begin{figure}[tbh]
\begin{center}
  \includegraphics[width=0.69\textwidth,clip=true]{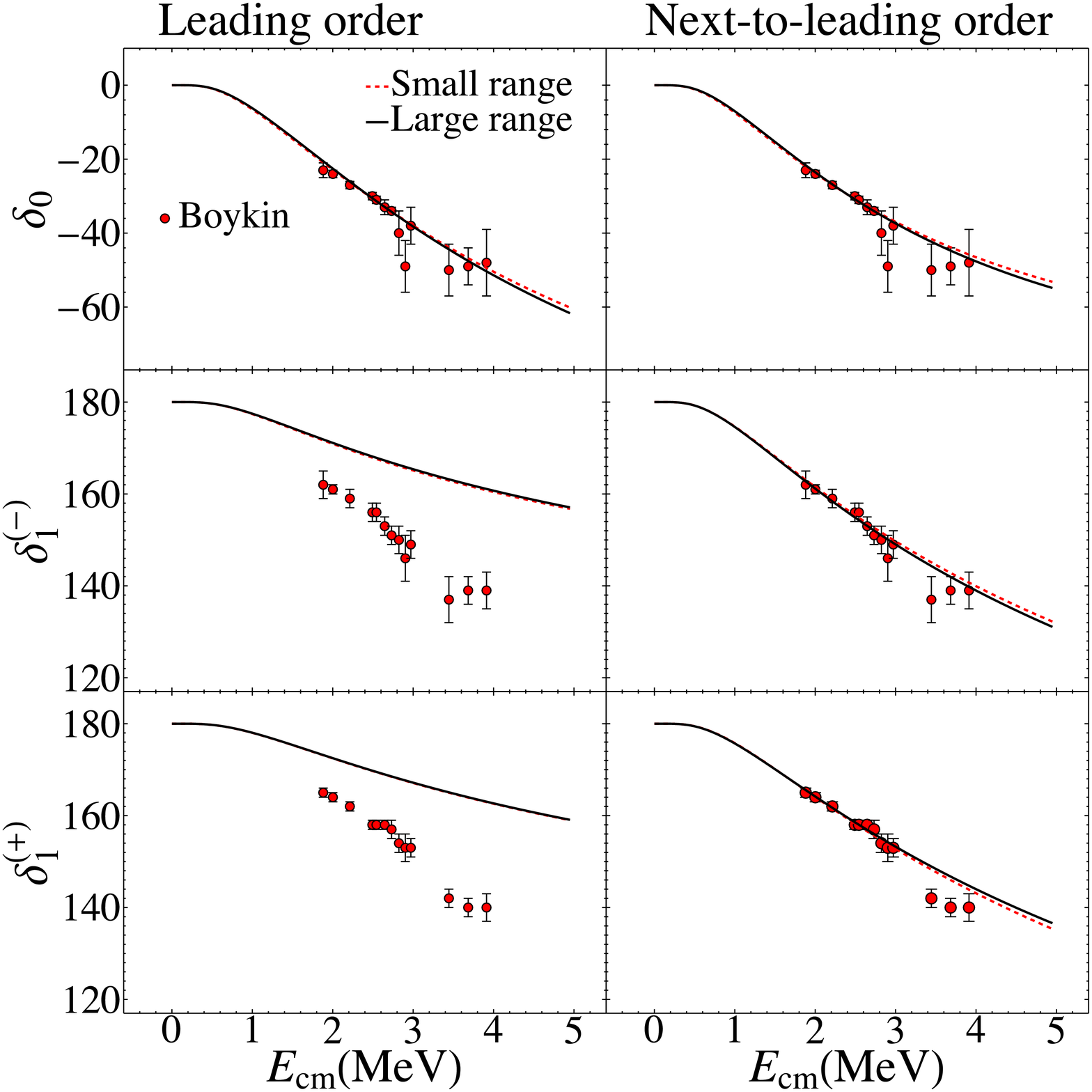}\end{center}
\caption{\protect $^{3}{\rm He}$-$\alpha$ scattering phase shifts at 
LO (left panel) and NLO (right panel), as functions of the c.m. energy 
$E_{\rm cm}$. The data is from Ref.~\cite{Boykin:1972}. The rest of the notation is the same as 
Fig.~\ref{fig:CaptureFits}. 
}
\label{fig:PhaseShiftFits}
\end{figure}

In Table ~\ref{table:fits}, we summarize our fits. The  parameters are 
self-consistent across the two different kinds of 
fits: Small range and  Large range.  This is reflected in the plots 
in Figs. ~\ref{fig:CaptureFits}, ~\ref{fig:RatioFits} and ~\ref{fig:PhaseShiftFits}.  The largest 
uncertainties were in the $s$-wave ERE parameters $a_0$ and $s_0$.  
Within the uncertainties, $a_0\sim \Lambda^2/Q^3$. The fitted $s_0$ values 
were slightly smaller than the expected $\sim 1/\Lambda^3$.  There is not 
a lot of difference in the LO and NLO capture cross sections at low energy,  
Fig.~\ref{fig:CaptureFits}.  This is consistent with a small $s_0$. The $d$-wave contributions are small, except at higher energies where  they, still of size NLO, bring 
the EFT numbers in better agreement with capture data. Even the Small range 
numbers that were fitted to below $500$ keV agrees with capture data around 
1 MeV. The $p$-wave ERE parameters are consistent with the power-counting as 
well, $\rho_1^{(\zeta)}\sim Q$, $\sigma_1^{(\zeta)}\sim 1/\Lambda$. The 
$p$-wave phase shifts are reproduced  at NLO once the shape parameter 
$\sigma_1^{(\zeta)}$ contributions are included, Fig~\ref{fig:PhaseShiftFits}.
The dimensionless two-body 
current couplings are of natural size, 
$L_\mathrm{E1}^{(\zeta)}\sim\mathcal O(1)$.

In Table~\ref{table:S34} we list $S_{34}(0)$ in EFT. The errors from the fits 
were propagated to the $S$-factor assuming a linear model as follows. For a 
function $f(\bm{r};\bm{\beta})$, where $\bm{r}$ is the independent variable 
and $\bm{\beta}$ the parameter set, we estimate the error as 
%----------- Equation
\begin{align}
  \delta f(\bm{r};\bm{\beta})&=\sqrt{
    \frac{\partial f(\bm{r};\bm{\beta})}{\partial\beta_i}
    \operatorname{COV}_{ij}
    \frac{\partial f(\bm{r};\bm{\beta})}{\partial\beta_j}}\,,
\end{align}
where $\mathrm{COV}_{ij}$ are the elements of the covariance matrix. 
We indicate a theory error of 30\% and 10\%, respectively, in our LO and NLO 
estimates. The central values of the EFT fits are compatible within the 
errors with other estimates ($0.593$ keV b from 
FMD~\cite{Neff:2010nm}, $0.59$ keV b from  
NCSM~\cite{Dohet-Eraly:2015ooa}, $(0.567\pm0.018\pm0.004)$ keV b 
from LUNA~\cite{LUNA}, $(0.595\pm0.018)$ keV b from 
Seattle~\cite{Seattle}, $(0.53\pm0.02\pm0.01)$ keV b from 
Weizmann~\cite{Weizmann}, $(0.57\pm0.04)$ keV b from 
ERNA~\cite{ERNA}, and $(0.554\pm0.020)$ keV b from Notre 
Dame~\cite{NotreDame}). 
%------------- Table 2 -----------------------
\begin{table}[htb]
\centering
\begin{ruledtabular}
  \begin{tabular}{ll}
       Fit & $S_{34}(0)$ (keV b) \\ \hline
       \begin{filecontents*}{table_S34_dwave.csv}
Row,Sfactor,Fit,EFT
Small range LO,0.582,0.011,0.194
Large range LO,0.561,0.007,0.187
Small range NLO,0.544,0.012,0.054
Large range NLO,0.558,0.008,0.056
\end{filecontents*}
\csvreader[head to column names, late after line=\\]{table_S34_dwave.csv}{}
{\Row &\Sfactor $\pm$ \Fit~(fit) $\pm$ \EFT~(EFT)}
     \end{tabular}
\end{ruledtabular}
\caption{$S_{34}$ at threshold. In the EFT results, the first error estimate 
is from the fits and the second is an estimated 30\% and 10\% error to the LO and NLO EFT results, respectively.
 Note that the EFT results are evaluated at 
$E=20\times 10^{-3}$ keV. 
}  
\label{table:S34}
\end{table}

\begin{figure}[tbh]
\begin{center}
  \includegraphics[width=0.69\textwidth,clip=true]{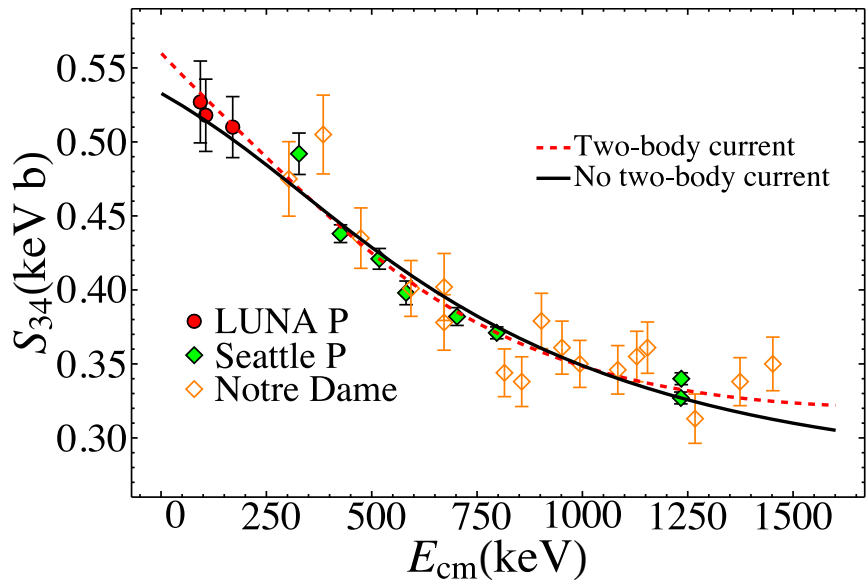}\end{center}
\caption{\protect $S_{34}$ as a function of the c.m. energy $E_\mathrm{cm}$. 
The fits include only prompt data from LUNA~\cite{LUNA}, 
Seattle~\cite{Seattle}, and Notre Dame~\cite{NotreDame}, with (red-dashed) 
and without (blue-solid) the two-body currents, to about 1200 keV. 
}
\label{fig:PotModel}
\end{figure}

Finally, we comment on potential model calculations that do not include the 
two-body current contributions. As mentioned, there is a large cancellation 
between the diagrams from Figs.~\ref{fig:CaptureTriangle} and 
\ref{fig:CaptureBubble}. In the absence of two-body currents this would 
require the wave function renormalization constant to be large. This, however, 
is not an issue since a small variation in $\rho_1^{(\zeta)}$ can cause the 
normalization constant to be huge. $[h^{(+)}]^2 \mathcal Z^{(+)}$ and  
$[h^{(-)}]^2 \mathcal Z^{(-)}$ have a pole at approximately $\rho_1^{(+)}= 
-47.4$ MeV and $\rho_1^{(-)}= -32.4$ MeV, respectively,  see 
Eq~(\ref{eq:wfrenormdef2}).  We explore this idea in Fig.~\ref{fig:PotModel} 
where $d$-wave contributions were not included to highlight the cancellation. 
For illustration we fit the prompt data from LUNA~\cite{LUNA}, 
Seattle~\cite{Seattle}, and Notre Dame~\cite{NotreDame} with and without 
the two-body current, to about 1200 keV. We also include some phase shift 
data. The results are similar. The curve without two-body currents used 
$\rho_1^{(+)}\approx -49$ MeV and $\rho_1^{(-)}\approx -36$ MeV, which are 
only slightly different than those in Table~\ref{table:fits} but lead to much 
larger wave function renormalization constant. The asymptotic normalization 
constants (ANC) were evaluated in potential models, see Ref.~\cite{Igamov:2009eh}. These 
are related to the wave function renormalization we calculate via 
%%----------- Equation --------------
\begin{align}
C_{1,\ \zeta}^2=\frac{[\gamma^{(\zeta)}]^2 \Gamma(2+k_C/\gamma^{(\zeta)})}{\pi} [h^{(\zeta)}]^2 \mathcal Z^{(\zeta)}\,.
\end{align}
Using the calculated ANCs~\cite{Igamov:2009eh} $C_{1,{}^2P_{3/2}}^2=23.3$ fm$^{-1}$ and $C_{1,{}^2P_{1/2}}^2=15.9$ fm$^{-1}$, we get $\rho_1^{(+)}= -52.3$ MeV and $\rho_1^{(-)}= -38.7$ MeV, respectively, in quite good agreement with the EFT numbers without two-body currents.

%%%------------- Figure 9 --------------------------------

%===========================================================
\section{Conclusions}\label{sec_summary}
%===========================================================
%\input{sec7-conclusion}
The capture cross section $\sigma$ and the related $S$-factor $S_{34}$ for 
\HeAlpha ~were calculated at low energies in halo EFT. E1 transitions from 
initial $s$- and $d$-wave states to the bound $p$-wave ground and excited 
states were considered. The Coulomb photons between the two clusters $^3$He 
and $\alpha$ were iterated to all orders in perturbation. 
The LO contributions to E1 transitions come from the initial $s$-wave  
Coulomb and short-range interactions, and from two-body 
currents. At NLO, $s$-wave shape parameter and pure Coulomb $d$-wave initial 
state interactions were included. 

We determine the low-energy parameters by a simultaneous fit of the halo 
EFT expressions to $S_{34}$ measurements, and $s$- and $p$-wave phase shifts. 
Two different fits were employed that gave very similar 
results. The fits corroborate the adopted power-counting where the $s$-wave 
scattering length and effective range both contribute at LO.  The contribution 
from the $s$-wave shape parameter is suppressed by a factor of $Q/\Lambda$ 
compared to LO. Pure Coulomb $d$-wave initial state interactions do not 
contribute significantly ($\lesssim 10$\%) up to $E_{\rm cm}\sim 500$~keV, 
and above that it behaves like a typical ($\lesssim 30$\%) NLO correction. 
We estimate a $10\%$ error in the EFT results from higher order NNLO 
contributions from initial state $d$-wave strong interactions and higher order 
two-body currents. 

Despite the apparently good description of data, reasonable agreement 
among the different fit schemes, and a converging pattern from LO to NLO, 
our fits use elastic scattering data mostly concentrated on the 
high-energy end of the validity range of the halo EFT. That causes certain 
instabilities in finding the optimal set of fitting parameters, since they 
should in principle be better fixed at lower energies where data are 
scarce or even non-existent. 
Nevertheless the halo EFT results for $S_{34}(0)$ are in good agreement with 
some model extrapolations of 
experimental measurements~\cite{LUNA,Seattle, Weizmann, ERNA, NotreDame}
and also recent ab 
initio calculations~\cite{Neff:2010nm,Dohet-Eraly:2015ooa}.

Moreover, we find with the current dataset that the contribution of two-body 
currents is strongly correlated to the wave function renormalization constants 
in a way that a decrease in one is compensated by an increase in the other. 
We showed the possibility of fitting phase shifts and capture data to our 
expressions without two-body currents, similar to what is done in potential 
model calculations. Experimental phase shifts at lower energies are likely 
to provide stronger constraints on the EFT parameters with great chance to 
disentangle the role of these terms. 
The planned TRIUMF 
experiment to measure  phase shifts down to c.m. energies of about 
500 keV~\cite{Davids:TRIUMF} would be useful to constrain the low-energy 
theory.
It would determine the wave function renormalization constants directly 
from elastic scattering data without resorting to fits to capture data. 
The expressions presented here can be directly applied to 
$^3$H$(\alpha,\gamma)^7$Li capture calculation where the main difference 
is the charge $Z_\psi=1$ of the $^3$H nucleus. 
The EFT parameters would have to be tuned to the $^3$H-$\alpha$ system. 
This work is under progress.

\acknowledgments
The authors thank D. R. Phillips  and X. Zhang for pointing out a mistake in one of our expressions, and other comments on the manuscript. 
G.R. thanks the Institute of Physics at  University of S\~{a}o Paulo, 
the Institute for Nuclear Theory at University of Washington,
the Kavli Institute of Theoretical Physics at University of California 
Santa Barbara, for kind hospitality during part of the completion of this 
research.
G.R. acknowledges partial support from the Joint Institute for Nuclear Physics and Applications at Oak Ridge National Laboratory and the Department of Physics, University of Tennessee during his sabbatical where part of this research was completed.
This work was partially supported by U.S. NSF grants PHY-1307453, 
PHY-1615092, Brazilian agency FAPESP-2012/50984-4, 
and Brazilian project INCT-FNA Proc. No. 464898/2014-5.

\bibliographystyle{apsrev4-1}

\end{document}